\documentclass{IEEEtran}
\usepackage{cite}
\usepackage{textcomp,nicefrac}
\usepackage{amsmath,amssymb,amsfonts}
\usepackage{algorithmic}
\usepackage{graphicx}
\usepackage[table]{xcolor}
\usepackage[separate-uncertainty=true]{siunitx}
\usepackage{glossaries}
\usepackage{lipsum}
\usepackage{booktabs}
\usepackage[hyphens]{url}
\usepackage{hyperref}
\usepackage[noabbrev,capitalise]{cleveref}
\usepackage{multirow}
\usepackage{tikz}
\usepackage{threeparttable}
\usepackage{tabularx}
\usepackage{caption}
\usepackage{subcaption}
\usepackage[export]{adjustbox}
\usepackage{orcidlink}
\usepackage{eqparbox}
\usepackage{soul}

\soulregister\gls7
\soulregister\cgls7
\soulregister\cglspl7
\soulregister\Gls7
\soulregister\cGls7
\soulregister\cGlspl7
\soulregister\acrshort7
\soulregister\acrlong7
\soulregister\Acrshort7
\soulregister\Acrlong7
\soulregister\Acrlong7
\soulregister\SI7
\soulregister\num7
\soulregister\cref7
\soulregister\Cref7
\soulregister\cite7
\soulregister\textcolor7

\definecolor{PULPred}{HTML}{A8322C}
\definecolor{PULPblue}{HTML}{1269B0}
\definecolor{PULPgreen}{HTML}{168638}
\definecolor{PULPorange}{HTML}{F29545}
\definecolor{PULPpurple}{HTML}{910569}

\glsenableentrycount

\newacronym[longplural=systems-on-chip]{soc}{SoC}{system-on-chip}
\newacronym{isa}{ISA}{instruction set architecture}
\newacronym{pulp}{PULP}{parallel ultra-low power}
\newacronym{odrg}{ODRG}{on-demand redundancy grouping}
\newacronym{tcls}{TCLS}{triple-core lockstep}
\newacronym{ecc}{ECC}{error correction codes}
\newacronym{dut}{DUT}{device under test}
\newacronym{wdt}{WDT}{watchdog timer}
\newacronym{seu}{SEU}{single event upset}
\newacronym{see}{SEE}{single event effect}
\newacronym{sefi}{SEFI}{single event functional interrupt}
\newacronym{set}{SET}{single event transient}
\newacronym{sel}{SEL}{single event latchup}
\newacronym{let}{LET}{linear energy transfer}
\newacronym{sdc}{SDC}{silent data corruption}
\newacronym{due}{DUE}{detectable unrecoverable error}
\newacronym{tid}{TID}{total ionizing dose}
\newacronym{rhbd}{RHBD}{radiation hardening by design}
\newacronym{ge}{GE}{gate equivalent}
\newacronym{udma}{$\mu$DMA}{I/O DMA}
\newacronym{gpio}{GPIO}{general purpose input/output}
\newacronym{dma}{DMA}{direct memory access}
\newacronym{tcdm}{TCDM}{tightly coupled data memory}
\newacronym{pdk}{PDK}{process desing kit}
\newacronym{pcb}{PCB}{printed circuit board}
\newacronym{sram}{SRAM}{static random-access memory}
\newacronym{tmr}{TMR}{triple modular redundancy}
\newacronym{dmr}{DMR}{dual modular redundancy}
\newacronym{secded}{SECDED}{single error correction, double error detection}
\newacronym{mftf}{MFTF}{mean fluence to failure}
\newacronym{mttf}{MTTF}{mean time to failure}
\newacronym{rpi}{RPi}{Raspberry Pi}
\newacronym{ff}{FF}{flip-flop}
\newacronym{fpga}{FPGA}{field-programmable gate array}

\DeclareRobustCommand{\hlA}[1]{{\sethlcolor{yellow}\hl{#1}}}
\DeclareRobustCommand{\hlB}[1]{{\sethlcolor{yellow}\hl{#1}}}

\renewcommand{\hlA}[1]{#1}
\renewcommand{\hlB}[1]{#1}

\newif\ifresultcolor
\newcommand{\result}[1]{{\ifresultcolor\textcolor{blue}{#1}\else#1\fi}}

\DeclareSIUnit\neutron{n}
\DeclareSIUnit\proton{p}
\DeclareSIUnit\error{error}
\newrobustcmd\B{\DeclareFontSeriesDefault[rm]{bf}{b}\bfseries}

\def\BibTeX{{\rm B\kern-.05em{\sc i\kern-.025em b}\kern-.08em
T\kern-.1667em\lower.7ex\hbox{E}\kern-.125emX}}
\begin{document}
\bstctlcite{IEEEexample:BSTcontrol}

\title{Trikarenos: Design and Experimental Characterization of a Fault-Tolerant 28nm RISC-V-based SoC}

\author{
Michael~Rogenmoser\textsuperscript{\textsection}\,\orcidlink{0000-0003-4622-4862},~\IEEEmembership{Graduate~Student~Member,~IEEE},
Philip~Wiese\textsuperscript{\textsection}\,\orcidlink{0009-0001-7214-2150},~\IEEEmembership{Graduate~Student~Member,~IEEE},
Bruno~Endres~Forlin\,\orcidlink{0000-0003-4822-1841},
Frank~K.~Gürkaynak\,\orcidlink{0000-0002-8476-554X},
Paolo~Rech\,\orcidlink{0000-0002-0821-1879},~\IEEEmembership{Senior~Member,~IEEE},
Alessandra~Menicucci\,\orcidlink{0000-0002-7064-6275},
Marco~Ottavi\,\orcidlink{0000-0002-5064-7342},~\IEEEmembership{Senior~Member,~IEEE},
Luca~Benini\,\orcidlink{0000-0001-8068-3806},~\IEEEmembership{Fellow,~IEEE}
\thanks{\textsuperscript{\textsection}Authors contributed equally to this work.}
\thanks{This work was supported in part through the TRISTAN project that has received funding from Chips Joint Undertaking (CHIPS-JU) under grant agreement nr. 101095947. CHIPS JU receives support from the European Union’s Horizon Europe’s research and innovation programme and Austria, Belgium, Bulgaria, Croatia, Cyprus, Czechia, Germany, Denmark, Estonia, Greece, Spain, Finland, France, Hungary, Ireland, Israel, Iceland, Italy, Lithuania, Luxembourg, Latvia, Malta, Netherlands, Norway, Poland, Portugal, Romania, Sweden, Slovenia, Slovakia, Turkey.}
\thanks{Michael Rogenmoser, Philip Wiese, and Frank K. Gürkaynak are with the Integrated Systems Laboratory at ETH Zurich, Switzerland (email:~\{michaero,~wiesep,~kgf\}@iis.ee.ethz.ch).}
\thanks{Bruno Endres Forlin is with the Computer Architecture for Embedded Systems of the University of Twente, the Netherlands (email:~b.endresforlin@utwente.nl).}
\thanks{Paolo Rech is with the Department of Industrial  Engineering of the University of Trento, Italy (email:~paolo.rech@unitn.it).}
\thanks{Alessandra Menicucci is with the Department of Space Engineering at the Delft University of Technology, the Netherlands (email:~a.menicucci@tudelft.nl).}
\thanks{Marco Ottavi is with the Computer Architecture for Embedded Systems of the University of Twente, the Netherlands and the Department of Electronic Engineering  at the University of Rome, Tor Vergata, Italy (email:~m.ottavi@utwente.nl).}
\thanks{Luca Benini is with the Integrated Systems Laboratory at ETH Zurich, Switzerland and the Department of Electrical, Electronic, and Information Engineering at the University of Bologna, Italy (email:~lbenini@iis.ee.ethz.ch).}
}

\maketitle

\begin{abstract}
RISC-V-based fault-tolerant \cgls{soc} designs are critical for the new generation of automotive and space \cgls{soc} architectures. However, reliability assessment requires characterization under controlled radiation doses to accurately quantify the fault tolerance of the fabricated designs.
This work analyzes the Trikarenos design, a \cgls{soc} implemented in TSMC 28nm, for \cgls{seu} vulnerability under atmospheric neutron and \SI{200}{\mega\eV} proton radiation, comparing these results to simulation-based fault injection.
All faults in \cgls{ecc} protected memory are corrected by a scrubber, showing an estimated cross-section per bit of up to \result{\SI{1.09e-14}{\centi\meter\squared\per\bit}}.
Furthermore, the \cgls{tcls} mechanism implemented in Trikarenos is validated and is shown to correct errors affecting a cross-section up to \result{\SI{3.23e-11}{\centi\meter\squared}}, with the remaining uncorrectable vulnerability below \result{\SI{5.36e-12}{\centi\meter\squared}}.
When augmenting the experimental analysis of fabricated chips with gate-level fault injection in simulation, \result{\SI{99.10}{\percent}} of injections into the \cgls{soc} produced correct results, while \result{\SI{100}{\percent}} of injections in the \cgls{tcls}-protected cores were handled correctly.
With \result{\SI{12.28}{\percent}} of all injected faults leading to a \cgls{tcls} recovery, this indicates an approximate effective flip-flop cross-section of up to \result{\SI{1.28e-14}{\centi\meter\squared\per FF}}.

\end{abstract}

\glsresetall


\section{Introduction}
\label{sec:introduction}
In the automotive and space domains, the reliability of semiconductor chips and \cglspl{soc} is critical.
The space sector faces challenges due to high radiation levels causing \cglspl{see}, which can disrupt the normal functioning of electronic components. Reliability is also a major intrinsic concern for the automotive industry since atmospheric neutrons an cause \cglspl{seu}, leading to crashes or unexpected behaviors. 
Various strategies are employed to mitigate these risks.
At the technology level, radiation-hardened technologies and specialized cells can enhance resilience \cite{narasimham_hysteresis-based_2012}.
Alternatively, architectural solutions such as \cgls{dmr}, \cgls{tmr} \cite{wilson_neutron_2023}, or \cgls{ecc} \cite{neale_neutron_2016} can be implemented to ensure a safe operation in critical applications. 

The RISC-V architecture is rapidly gaining traction in safety-critical applications, thanks to the openness of the \cgls{isa} and the availability of open processor IPs~\cite{di_mascio_open-source_2021}. As a key advantage, the flexibility of RISC-V allows customization and improvement of the \cgls{isa} and the underlying hardware, tuning them to meet specific (reliability) requirements.
Consequently, the RISC-V ecosystem is increasingly adopted in applications where system reliability is non-negotiable~\cite{walsemann_strv_2023}.

Within this context, previous investigations have tested the reliability of various RISC-V processors, including processors with architectural protection, under neutron and proton radiation~\cite{de_oliveira_evaluating_2020,wilson_neutron_2023,de_oliveira_evaluating_2023,santos_hybrid_2024}.
However, these investigations have been limited to \cgls{fpga} implementations of these processors, leaving the configuration memory vulnerable to \cglspl{seu}.

\begin{figure*}[t]
    \centering
    \begin{subfigure}[b]{0.38\linewidth}
        \centering
        \includegraphics[width=0.95\textwidth]{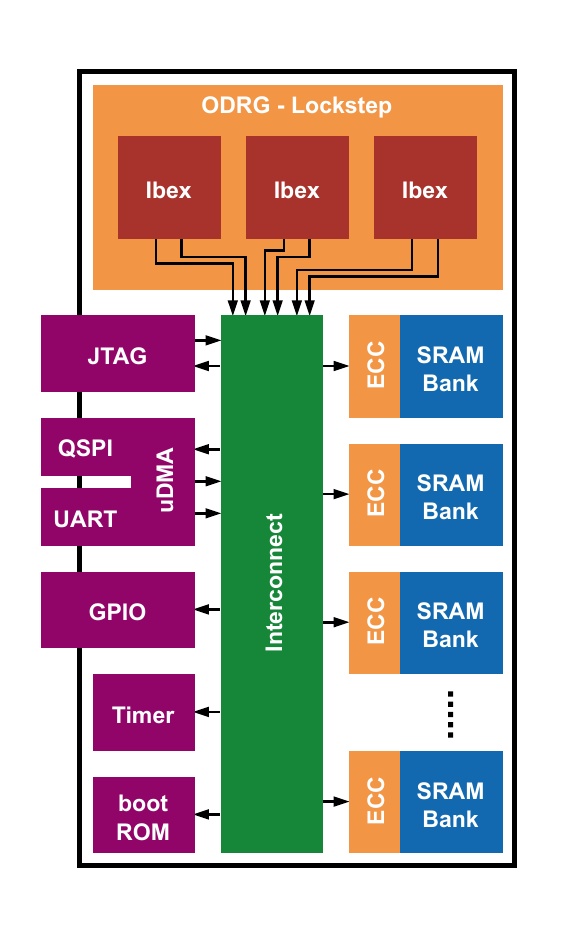}
        \caption{Block diagram of Trikarenos \acrshort{soc}, containing three Ibex cores for fault-tolerant operation, \acrshort{ecc}-protected memory, and peripherals, connected using a low latency interconnect.}
        \label{fig:trik-block}
    \end{subfigure}
    \hfill
    \begin{subfigure}[b]{0.6\linewidth}
    \begin{subfigure}{\linewidth}
        \centering
        \includegraphics[width=0.95\textwidth]{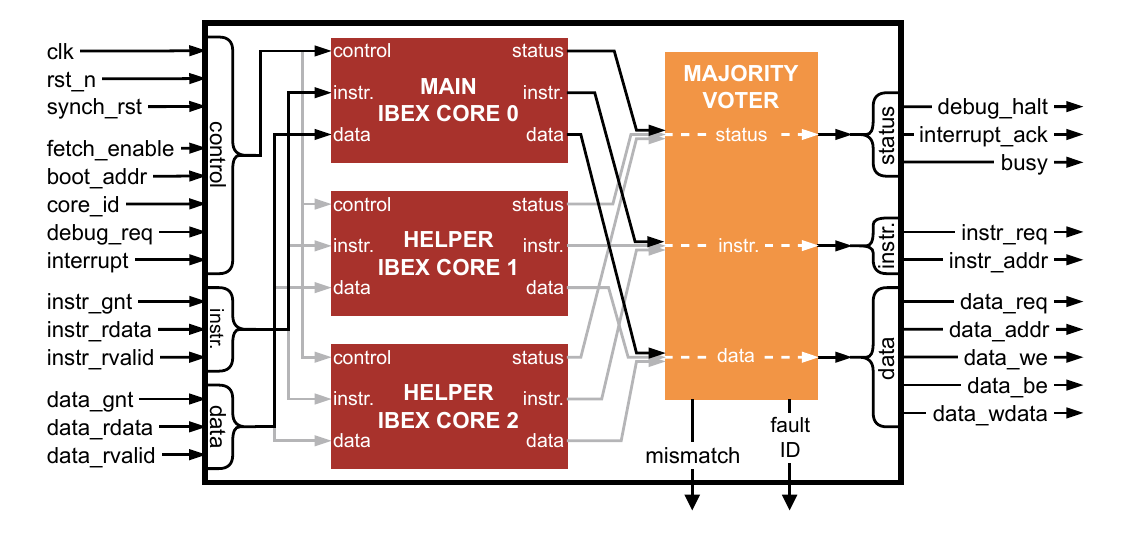}
        \caption{Diagram of the \acrfull{tcls} mechanism with three Ibex cores. The three core's input signals, split into control, instruction, and data, are identical, with the output signals voted, with additional information indicating an error and the responsible core.}
        \label{fig:ODRG}
    \end{subfigure}
    \vspace{0.1cm}
    \\
    \begin{subfigure}{\linewidth}
        \centering
        \includegraphics[width=0.95\textwidth]{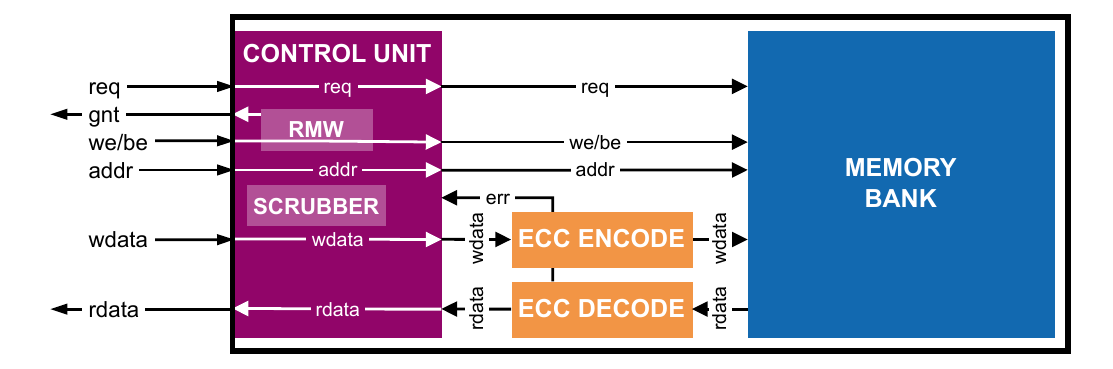}
        \caption{Diagram of the \acrshort{sram} error correction scheme. Each 32-bit word is stored in 39~bits with a \acrlong{secded} Hsiao code. A scrubber continuously checks the memory for errors, ensuring real-time correction without system interruptions.}
        \label{fig:ecc_mem}
    \end{subfigure}
    \end{subfigure}
    \caption{Trikarenos \acrshort{soc} architecture, including (a) the overall system block diagram, (b) the \acrshort{tcls} mechanism for core fault tolerance, and (c) the \acrshort{sram} error correction and scrubbing system. The design is optimized for fault tolerance in automotive and space applications.}
    \label{fig:trik-arch}
\end{figure*}

Our paper focuses on the experimental evaluation of \textit{Trikarenos}~\cite{rogenmoser_trikarenos_2023}, a \cgls{soc} designed to address the reliability requirements of both automotive and space applications.
Trikarenos leverages the flexibility of the RISC-V architecture, integrating a \cgls{tcls} system alongside \cgls{ecc} and scrubbing in memory, implemented in TSMC 28nm technology.
This work investigates the architectural protection features of Trikarenos under neutron and proton radiation as well as gate-level simulations, evaluating its resilience for automotive and space contexts.
Specifically, our contributions include:
\begin{itemize}
    \item The measurement of \cgls{sram} vulnerability within Trikarenos, focusing on the effectiveness of \cgls{ecc} protection and scrubbing techniques.
    \item An evaluation of the \cgls{tcls} mechanism, demonstrating its capacity to maintain the functionality of the processor in the event of a fault.
    \item A performance assessment of Trikarenos in various radiation environments, including an atmospheric setting dominated by neutrons and an orbital environment characterized by protons.
    \item Evaluation of Trikarenos' vulnerability to \cglspl{seu} using simulation-based fault injection, comparing to experimental results.
\end{itemize}

\section{\hlA{Background}}

\subsection{\hlA{Radiation Hardening Techniques}}
\hlA{Radiation hardening employs a range of strategies to enhance the resilience of systems against radiation-induced faults, particularly in demanding environments such as space.}
\hlB{These strategies can be broadly categorized into software-based fault tolerance techniques, radiation-hardened technologies, and design-level mitigation approaches~\cite{ecss_secretariat_space_2023, rogenmoser_hybrid_2025}.}
\hlB{Trikarenos applies \cgls{rhbd}}\hlA{ techniques at the physical layout, circuit architecture, or system levels to mitigate the effects of radiation on integrated circuits.
One widely used \cgls{rhbd} technique is \cgls{tmr}, which provides fault tolerance by replicating hardware modules three times and using a majority voter to determine the correct output. 
While highly effective at mitigating faults, \cgls{tmr} imposes overheads in terms of power and area, making it best suited for applications requiring high reliability.
\cGls{ecc} offer another effective approach, particularly for protecting data stored in memories and registers. 
By appending parity bits to the primary data, \cgls{ecc} enables the detection and correction of single-bit errors and, in certain implementations, multi-bit errors. 
Compared to replication-based methods, \cgls{ecc} provides efficient static data protection with minimal resource overhead.
By combining these techniques, radiation-hardened systems can meet the high-reliability requirements of space missions and other high-radiation environments while minimizing penalties in performance and cost.}

\subsection{\hlA{RISC-V Processors}}

\hlA{
In recent years, the adoption of RISC-V has globally advanced, including the space domain driven by mandates in Europe and high-profile projects like NASA's High-Performance Spaceflight Computing (HPSC), which incorporates RISC-V cores designed by SiFive~\cite{leibson_nasa_2023}.
Unlike proprietary \cglspl{isa} such as x86 and ARM, RISC-V offers a completely open and extensible architecture, enabling designers and researchers to implement custom modifications and specialized designs without requiring licensing fees.
In comparison to other \cglspl{isa} like SPARC and PowerPC, RISC-V delivers a modern, modular architecture with extensibility at its core. 
This design philosophy allows for the inclusion of custom instructions tailored to specialized workloads, addressing the needs of evolving computational demands, as well as custom modification to the designs, such as those targeting reliability.}

\hlA{One example of the practical potential of RISC-V is PULPissimo~\cite{schiavone_quentin_2018}, an open-source microcontroller architecture developed at ETH Zurich. 
It is designed for flexibility and accessibility and offers a configurable processing core, low-latency interconnect, and energy-efficient peripherals.
Its modular design enables the integration of custom components, making it adaptable to a wide range of applications and specialized use cases.}

\subsection{\hlA{Reliable Microprocessors}}

\hlA{A variety of radiation-tolerant \cglspl{soc} are available commercially. Processors such as the RAD750~\cite{rea_powerpc_2005} and newer RAD5500~\cite{berger_quad-core_2015} feature a PowerPC architecture with a proven track record operating in space. Frontgrade Gaisler's GR716A~\cite{johansson_rad-hard_2016} and GR716B offer a radiation-tolerant microcontroller platform using the SPARC \cgls{isa}, with newer high performance designs such as the GR765 and NOEL-V \cgls{soc}~\cite{andersson_gaisler_2022} offering RISC-V cores.

While these systems often report tolerance to radiation, this reporting is limited to \cgls{tid} and \cgls{sel} \cgls{let}. They do not report the tolerance to \cglspl{seu} or the cross-section thereof under proton or neutron radiation, making it difficult to appropriately compare to the results in this work.}

\section{SoC Architecture Description} \label{sec:architecture}

The Trikarenos architecture, presented in~\cite{rogenmoser_trikarenos_2023}, implements an adaptation of PULPissimo, a RISC-V-based microcontroller design.
As shown in \Cref{fig:trik-block}, Trikarenos includes three physically separated Ibex cores~\cite{schiavone_slow_2017} connected to a low latency interconnect, exposing \SI{256}{\kibi\byte} of \cgls{sram} memory.
Furthermore, the \cgls{soc} features a boot memory, timers, peripherals such as \cglspl{gpio}, UART, and QSPI with \cgls{dma}, and a JTAG debug unit for programming.

\subsection{Processing Core Reliability}

To boost tolerance to \cglspl{seu} at an architectural level, Trikarenos' cores are augmented with \cgls{odrg}~\cite{rogenmoser_hybrid_2025}, a configurable \cgls{tcls} mechanism.
In locked mode, the three cores' outputs, including instruction fetch and data request ports, are voted.
The cores' inputs, including the instruction and data response and interrupt requests, are identical.
This design, illustrated in \Cref{fig:ODRG}, ensures the cores all receive the same data, and any error propagating outside the cores is detected and corrected.
Once an error is detected at the interface, a recovery routine is started, ensuring the internal state of the cores is saved to memory, corrected through the voters, and rewritten to ensure an identical starting point for lockstep operation. \hlA{The \cgls{tcls} recovery mechanism adds up to 700 cycles overhead for the recovery routine in case of a mismatch is detected within the cores. In case a critical code section is being executed, this recovery can be delayed.}
Functionally, these three cores operate as a single core within the system.

\hlA{As the \cgls{tcls} protection is applied outside the processor cores at their interface, only individual voter cells are added in these paths, with the remaining logic not affecting the critical path. Even if the critical path for the design is through this interface, the impact on frequency remains minimal and was not observed to affect the target frequency for Trikarenos.}

\hlA{The system allows the three cores to operate independently in parallel for increased performance or in locked reliability mode, where the cores work together to detect and correct mismatches for fault tolerance. 
The mode switching is managed at runtime via system control registers.
This investigation focuses on the locked reliability mode, focusing on its capability of detecting and correcting any mismatch in the cores' operation.}

\subsection{Memory Reliability}

The on-chip \cgls{sram} memory, occupying most of the design area, is split into eight word-interleaved 32-bit memory banks, enabling low contention and supporting byte-wise writes. For reliability, each 32-bit word is stored in 39~bits with a \cgls{secded} Hsiao code, resulting in effectively \SI{2555904}{bits} for the \SI{256}{\kilo\byte} memory.

As shown in \Cref{fig:ecc_mem}, a control unit is connected to the encoders and decoders for each memory bank, supporting an efficient read-modify-write architecture for sub-word writes, immediately accepting the operation to allow the system to continue, delaying the subsequent access in case the bank is accessed in the following cycle to re-encode the stored data properly. \hlA{With the word-interleaved banks in Trikarenos, repeated access to the same bank is reduced significantly, and as only a very small amount of accesses to memory are writes of less than} \SI{32}{\bit}, \hlA{the performance impact is negligible. }

\hlA{The \cgls{ecc} encoding and decoding logic adds some additional latency on the data path before the memory, which could have an impact on the maximum achievable frequency. However, the use of an efficient \cgls{ecc} such as the Hsiao code limits the logic levels added to the path, and as other signals, such as handshake signals and address, have longer critical paths than the logic in the dataplane, this increase does not affect the achievable frequency.}

To ensure that latent errors do not accumulate within the memory, which would result in uncorrectable data, each memory bank is equipped with a scrubber, continuously scanning the memory bank contents at a configurable rate. Further, the scrubber delays its memory bank access if the system requests data from the memory bank in the same cycle\hlA{, thus having no impact on the system's performance}. In case the scrubber detects an error, it is directly corrected, and the correction \hlA{event} is logged.

\subsection{SoC Design}

\begin{figure}[t]
    \centering
    \includegraphics[width=0.7\columnwidth]{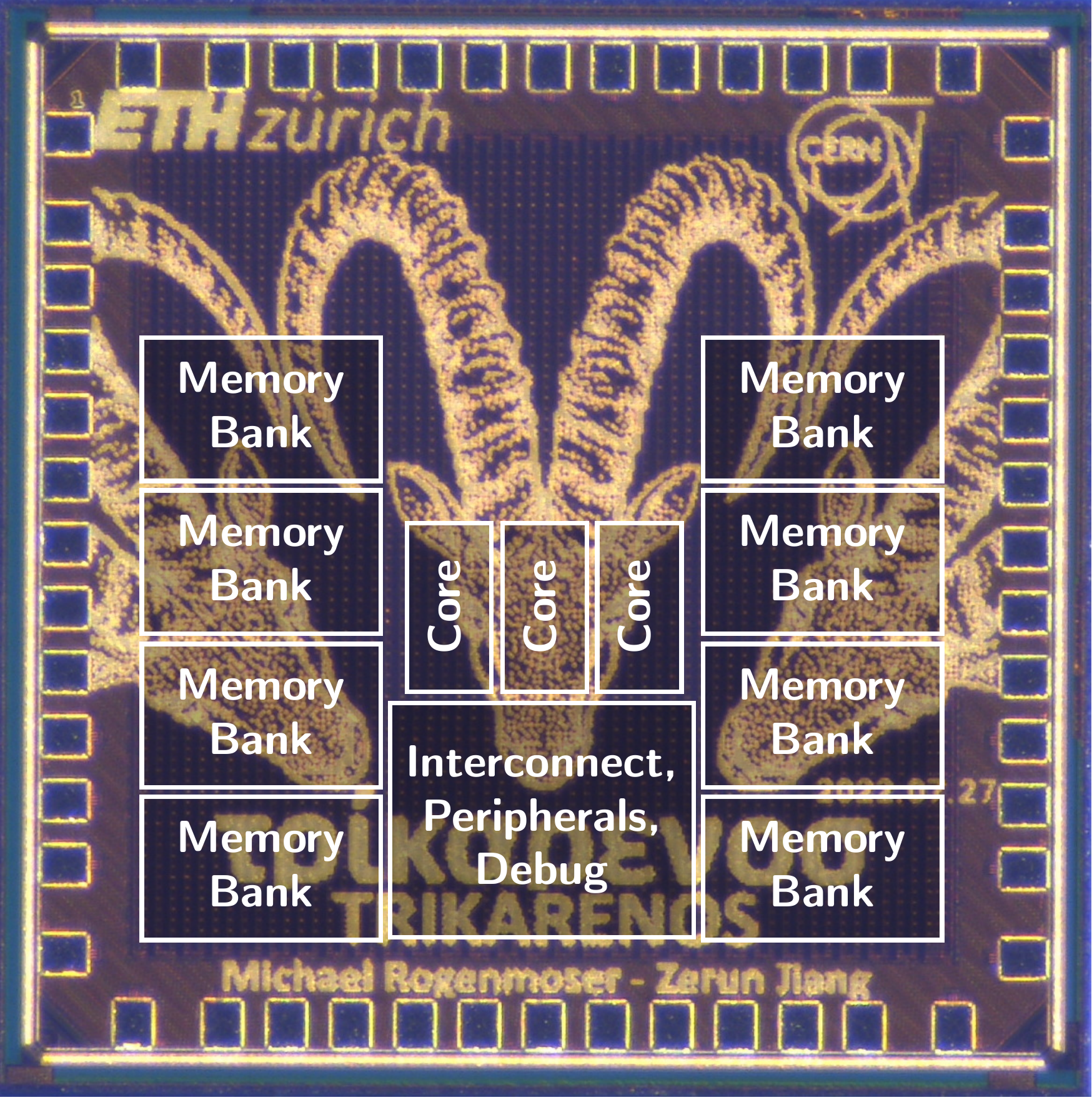}
    \caption{Annotated die shot of Trikarenos, highlighting the spatial distribution of the three Ibex cores, memory banks, interconnect, and key peripherals within the \SI{2}{\milli\meter\squared} die.}
    \label{fig:trik-die}
\end{figure}

\hlA{Along with the protected core and memory banks, Trikarenos' \cgls{soc} features a low-latency interconnect between the core, memory banks, and additional \cgls{soc} components such as debug module and peripherals including UART, SPI, and GPIOs. While the interconnect is critical to operation, it is far smaller in the area it occupies within the system. In an integrated system, the peripherals and debug module are used far less extensively than the processing core, however they have the ability to both send requests into and receive requests from other components through the interconnect. Interconnect, debug module and other peripherals are not modified from the base PULPissimo design to harden against radiation effects.}

To track the \cgls{soc}'s status, the architecture features a variety of telemetry registers counting errors in the system. This allows efficient readout of errors in the memory banks detected on access and detected by the scrubber, both for correctable and uncorrectable faults.
\hlA{Along with the interconnect, peripherals, and debug module, the \cgls{tcls} voting logic and \cgls{ecc} encoders and decoders are not hardened for reliability. 
Faults in these areas can lead to functional errors such as timeouts, exceptions, or incorrect results.
However, the voting logic and \cgls{ecc} encoders and decoders represent a relatively small fraction of the overall area compared to the cores and memory banks.}

\subsection{\hlA{Implementation}}
Trikarenos is implemented in TSMC 28HPC+, a 28nm bulk process that has been shown to be tolerant to \cgls{tid} effects~\cite{borghello_total_2023}. \hlA{Furthermore, TSMC 28HPC+ offers a good balance between performance, power efficiency, and accessibility, making it a practical choice for developing radiation-tolerant designs without requiring custom radiation-hardened processes.} The design was implemented relying on standard components and cells without hardened registers, adding physical separation between the processor cores but not adding any additional consideration for clock and reset tree hardening.
The implementation targets operating frequencies up to \SI{250}{\mega\hertz} (at \SI{0.9}{\volt}), occupying \SI{0.17}{\milli\meter\squared} for logic and \SI{0.56}{\milli\meter\squared} for \cgls{sram} in the \SI{2}{\milli\meter\squared} design~\cite{rogenmoser_trikarenos_2023} shown in \Cref{fig:trik-die}. In the experimental setup, Trikarenos operates at \SI{125}{\mega\hertz} and \SI{0.9}{\volt}.

\begin{figure}[ht]
    \centering
    \includegraphics[width=1\columnwidth]{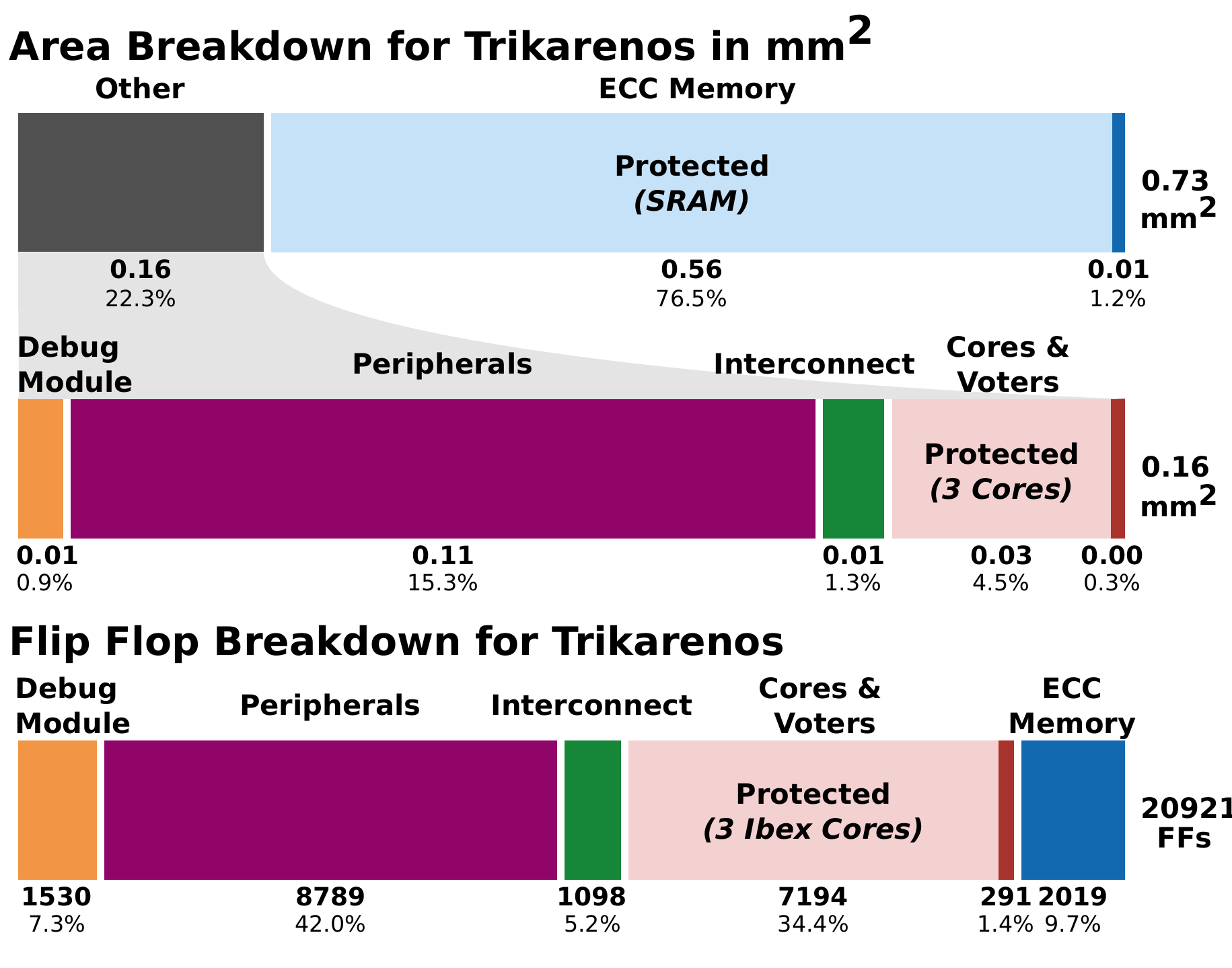}
    \caption{\hlA{Area and \cgls{ff} breakdown for the Trikarenos \cgls{soc}. The three Ibex cores and memory banks are protected, while the remaining components remain unprotected.}}
    \label{fig:trikarenos_area-ff}
\end{figure}

\hlA{\Cref{fig:trikarenos_area-ff} provides a detailed breakdown of the area and \cgls{ff} distribution across key components of the design, including the cores, interconnect, peripherals, memory, and debug modules.}
To provide context for the errors seen in experiments, one core contains \result{\SI{2398}{\cglspl{ff}}}, whereas the \cgls{soc} contains \result{\SI{20921}{\cglspl{ff}}}.

\section{Evaluation Methodology} \label{sec:methodology}

Two methods were used to evaluate the Trikarenos \cgls{soc}.
Primarily, beam experiments were used to determine the vulnerable cross-section in different parts of the design.
Fault injection in a simulation environment was also performed, relating the experimental cross-section to \cgls{ff} vulnerability.

\subsection{Testing Facilities}
We conducted experiments at the ChipIR and HollandPTC facilities to assess the sensitivity of Trikarenos to neutron and proton radiation.
The ChipIR beamline, located at the Rutherford Appleton Laboratories in the United Kingdom, is designed to replicate the energy spectrum of atmospheric neutrons~\cite{cazzaniga_progress_2018}. 
On the other hand, HollandPTC, located in Delft, The Netherlands, is equipped with a fixed horizontal proton beamline, delivering a beam with nominal energies ranging from \SI{70}{\mega\eV} up to \SI{240}{\mega\eV}. Its Cyclotron can deliver nominal beam currents from \SI{1}{\nano\ampere} up to \SI{800}{\nano\ampere} and a maximum flux of \SI{1.13E+09}{\proton\per\centi\meter\squared\per\second}~\cite{rovituso_characterisation_2023}.

\subsection{Error model}

Our experiments focus on analyzing \cglspl{see} in the RISC-V-based Trikarenos \cgls{soc}.
For the TSMC 28nm technology, other research has already investigated the \cgls{tid} effects~\cite{borghello_total_2023}, showing extreme tolerance levels.

We investigate the error mechanisms in Table~\ref{tab:error_types}, which are traceable with the \cgls{soc} telemetry registers and given test applications.

\begin{table}[ht]
    \centering
    \caption{Observable error types and classification}
    \begin{tabular}{@{}ll@{}}\toprule
        Observed Error & Class \\\midrule
        Single \cgls{sram} Errors (scrubber) & Correctable \\
        Multiple \cgls{sram} \hlB{Errors} (on access \& scrubber) & Functional Error \\
        \cGls{tcls} Events & Correctable \\
        Miscalculations (\cgls{sdc}) & Functional Error \\
        System Crashes & Functional Error \\
        \cGls{sel} & Total Loss \\\bottomrule
    \end{tabular}
    \label{tab:error_types}
\end{table}

While most error sources can be traced back to \cglspl{seu} or multiple/accumulation thereof, system crashes can have different causes. They can be traced back to a \cgls{sefi} or a \cgls{seu} affecting an unprotected, critical component, such as the system's interconnect. While some \cgls{gpio} signals and routines are designed to assist in differentiating these faults, exact cause identification remains difficult.

Finally, most data corruption will be detected by one of the two mitigation mechanisms.
\cGls{ecc} protection in the \cgls{sram} memory will catch an error affecting stored data, and the \cgls{tcls} mechanism will detect data corruption in the cores.
For a particle strike in the cores, it is more than likely that only a single core will be affected due to their separation, thus resulting in different outputs from the cores and an error in the voter while storing the result back in memory.

\subsection{Experimental Setup}
\begin{figure}[t]
    \centering
    \begin{subfigure}[b]{\columnwidth}
        \centering
        \includegraphics[width=0.9\columnwidth]{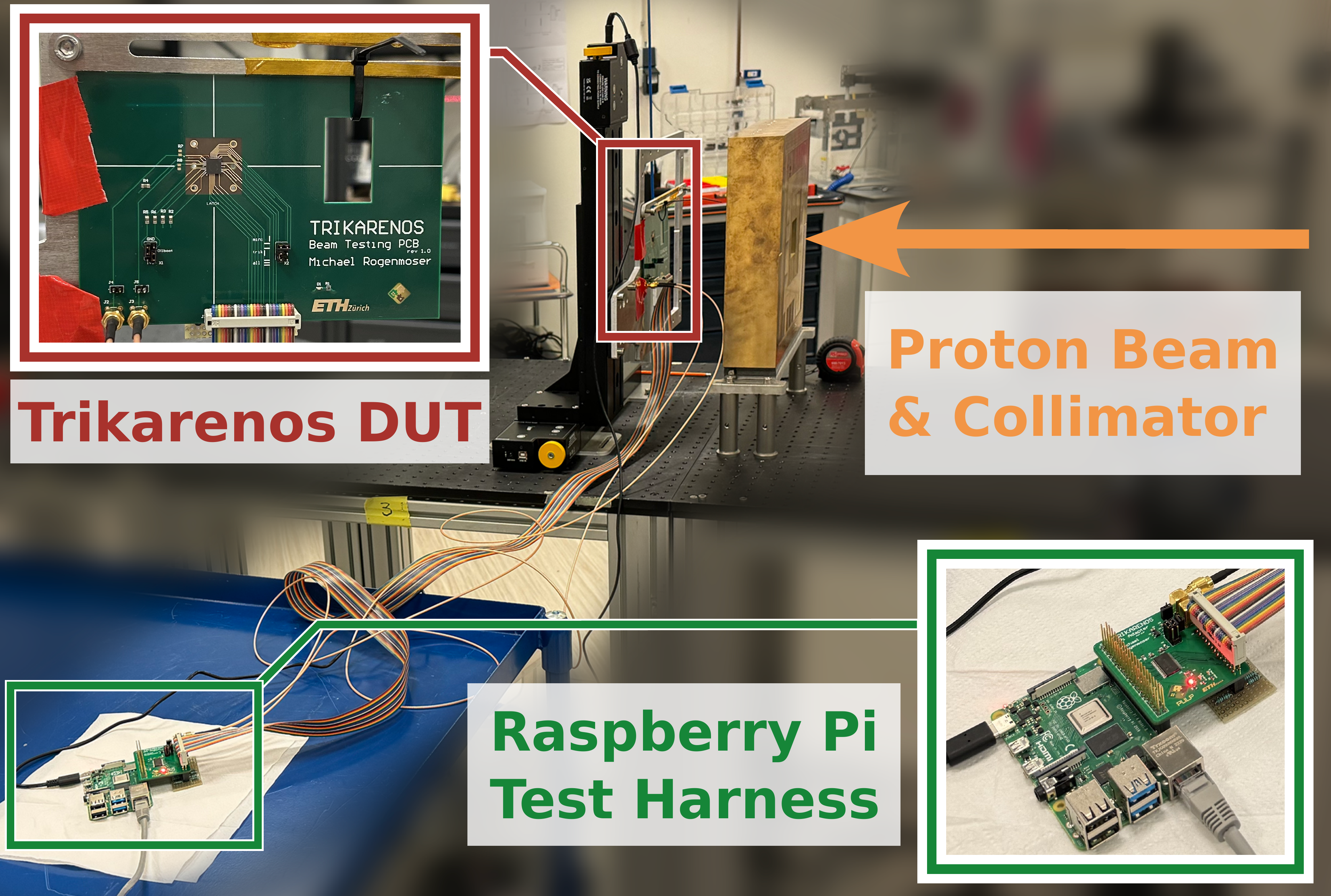}
        \vspace{0.2cm}
    \end{subfigure}
    \begin{subfigure}[b]{\columnwidth}
        \centering
        \includegraphics[width=0.9\columnwidth]{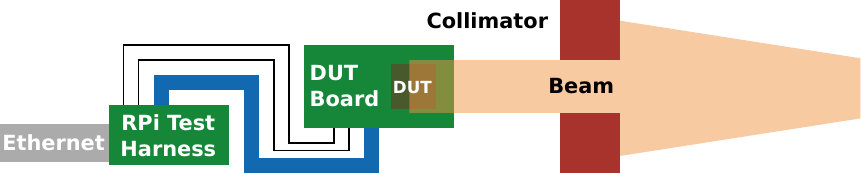}
    \end{subfigure}
    \caption{Annotated image of testing setup at HollandPTC (top) and schematic of the testing setup (bottom), showing the proton beam, the \acrfull{dut}, and the test harness.}
    \label{fig:test-setup}
\end{figure}

The test setup, shown in \Cref{fig:test-setup}, comprised a dedicated testing \cgls{pcb} containing the \cgls{dut} and a test harness, a monitoring device with an interface board. 
A \cgls{rpi} single-board computer served as the primary control and monitoring system for the entire test suite.
The presence of a Linux operating system on the Raspberry Pi enabled the development of flexible and modular software in Python, which facilitated comprehensive monitoring and control capabilities.
The interface board supplied the necessary \SI{1.8}{V} I/O and \SI{0.9}{V} core voltages for the Trikarenos \cgls{soc}, in addition to incorporating a level shifter, dual clock generation circuits, and several status LEDs for real-time feedback and diagnostics.
This setup controlled the power supply to the \cgls{dut}, managed binary loading into the \cgls{sram} via JTAG, and monitored \cgls{gpio} pins while interfacing with the \cgls{dut} via JTAG and UART for data acquisition.

The test application running on the \cgls{dut} was carefully designed to emulate a realistic application and maximize the propagation probability and observability of bit-flips inside the cores and memory.
To emulate a realistic application, Trikarenos executed the Coremark benchmark, implementing a variety of workloads to test a processor core.
The final version for the proton tests augmented this with a section continuously executing simple \textit{not} operations on the processor core's registers filled with a checkerboard pattern and periodically saving these values to memory.
Saving the register contents uses the \cgls{tcls} voters to detect any errors.
This maximized the exposure time of the vulnerable components, allowing for a possible mismatch between the cores in \cgls{tcls} to trigger an error.

For testing, the \cgls{dut} application facilitated real-time error monitoring by periodically transmitting the status of memory-mapped registers that count errors through the UART and JTAG interfaces.
\hlA{These registers include error counters that log recoverable \cgls{tcls} events detected by majority voting, \cgls{ecc} access errors that track memory read/write operations with invalid \cgls{ecc} bits, and scrubber counters that distinguish between corrected and uncorrectable errors in the memory.}
The test harness acquired this transmitted data, additionally monitoring other \cgls{gpio} signals designed to indicate other system failures.
Finally, in the event of critical, non-recoverable errors, the monitoring system would halt the cores and dump all relevant registers and memory areas for subsequent analysis.
This comprehensive approach, combined with multiple redundant recovery mechanisms and extensive logging strategies, equipped the setup for effective long-term and unsupervised radiation testing.

Data integrity and logging were critical concerns addressed in our setup.
All test data was logged locally on NAND flash memory using an SD card.
To ensure a proper backup in case of SD card corruption, the data, and \cgls{rpi}'s system logs to assist in tracing errors were transmitted to a computer outside the beam room over ethernet for later processing.
To detect latch-up events, a current monitor and a current-limited power switch were added to the test harness to prevent permanent damage to the \cgls{dut}.

\begin{figure*}[t!]
    \centering
    \includegraphics[width=0.975\linewidth]{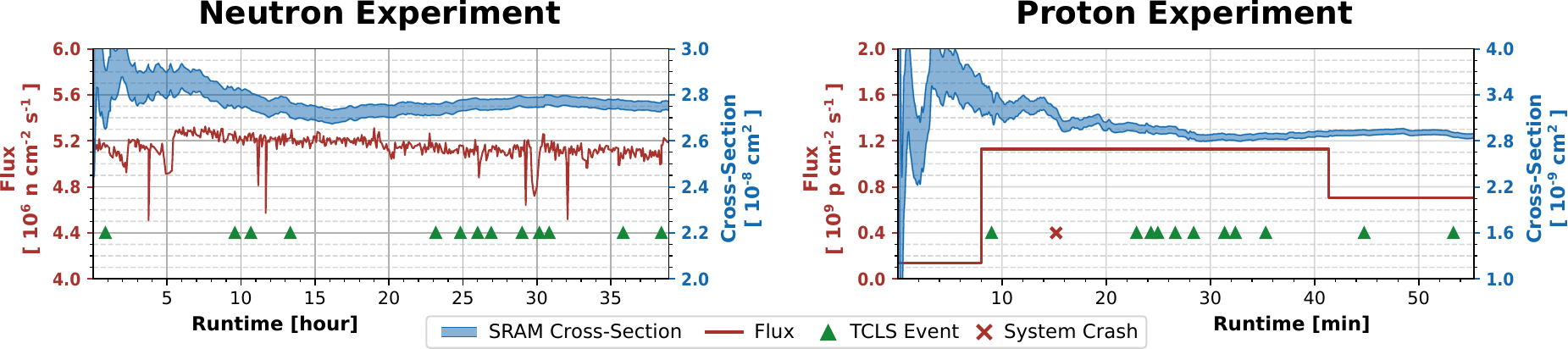}
    \caption{Visualization of induced errors and flux rates during both radiation tests of Trikarenos. We show the cross-sections, flux rates, \cgls{tcls} events, and system crashes over time. We use an atmospheric-like neutron beam~\cite{cazzaniga_progress_2018} and \SI{200}{\mega\eV} protons. For the proton tests, no \cgls{tid}-related degradation was observed.}
    \label{fig:results}
\end{figure*}

\subsection{Simulation-based Fault Injection}
To determine the vulnerability of the design to \cglspl{seu} in a more targeted manner, we performed extensive fault injection simulations on the post-layout gate-level netlist. 
The fault injection simulations were performed using \textit{Siemens QuestaSim 2023.4} with the help of custom scripts\footnote{\url{https://github.com/pulp-platform/InjectaFault/}}.
These scripts initially launch a normal simulation of the design without any modification, using this as a golden result to compare to, saving an initial checkpoint after setup has been completed, and storing the final state of all registers as a reference.
Once this reference is available, the scripts re-run the final part of the simulation from the checkpoint, selecting a random signal and flipping it at a random timestep in a given window using the \texttt{force} command.
The final result is then compared to the golden reference, differentiating correct results, functional errors, and timeouts, also keeping track of latent errors still in the design's registers.
These simulation runs are then repeated with different randomization seeds, storing the results for different tests.

During the simulation campaign, single bit-flips were injected into all registers of the \cgls{soc}.
\hlA{Bit-flips in registers or \cglspl{ff} are intended to represent \cglspl{seu}, both caused by direct flips within the registers or \cglspl{set} sampled by a \cgls{ff} disrupting functionality.}
Trikarenos executed the Coremark benchmark, followed by 100 iterations of \textit{not} operations on the processor core's registers, representing a single iteration of the program loop run during the beam experiments.
The injection scripts were adapted to differentiate the correct results further, checking for recoveries with \cgls{tcls} but ensuring the Coremark check returned a correct result.
Latent errors were filtered into two categories: some errors remain only in the originally injected register, not propagating through the system, indicating this register has limited functionality or the component injected is not used.
While propagating errors may indicate future issues with the device, the result and functionality were not impacted.
Functional errors were also differentiated, separating a simulation timeout, indicating an internal issue or infinite loop, an unexpected system exception, or an incorrect result from the test.

\section{Experimental Beam Results} \label{sec:results}

Throughout all our tests, we did not observe any uncorrectable \cgls{sram} errors, miscalculations, or \cglspl{sel} indicated by a change in power consumption.
Therefore, the results below only differentiate correctable \cgls{sram} errors, correctable \cgls{tcls} events, and uncorrectable errors that result in a system crash.
Because some errors are rare or possibly even zero, we calculate upper and lower estimates of the cross-section assuming a Poisson distribution and use the relationship between the cumulative distribution function of the Possion and the chi-squared distribution~\cite{johnson_poisson_2005}.
To calculate these estimates $\sigma$ we use
\begin{equation}
    \frac{1}{2 F} \chi^2\left( \frac{\alpha}{2}, 2 N \right) < \sigma < \frac{1}{2 F} \chi^2\left( 1-\frac{\alpha}{2}, 2 (N+1) \right)
    \label{equ:crosssection}
\end{equation}
where $\chi^2(p, n)$ is the quantile function of the chi-squared distribution with $n$ degrees of freedom, $\alpha$ describes the $100(1-\alpha)$ percent confidence interval, $F$ is the total exposed fluence, and $N$ is the number of detected errors.
All calculations below use a \SI{95}{\percent} confidence interval.

To differentiate between errors attributed to the test setup and those inherent to the \cgls{dut}, we conducted a detailed analysis of the system logs from the \cgls{rpi}.
Our evaluation identified that most errors originating from the setup involved SD card-related timeouts and kernel segmentation faults. 

\subsection{Neutron Tests}

At the ChipIR facility~\cite{cazzaniga_progress_2018}, Trikarenos was exposed to a total fluence of \result{\SI{6.88e11}{\neutron\per\centi\meter\squared}}.
While active, the beam produced an average neutron flux of \result{\SI{4.90e06}{\neutron\per\centi\meter\squared\per\second}} with an energy spectrum representing that found for atmospheric neutrons at ground level up to \SI{800}{\mega\eV}.
The measurements are shown in \Cref{fig:results}, and the errors and estimated cross-sections are summarized in \Cref{tab:neutrons}.

\begin{table}[ht]
\centering
\begin{threeparttable}
    \caption{Cross-section results of Trikarenos under neutron beam for a fluence of \result{\SI{6.88e11}{neutrons\per\centi\meter\squared}}}
    \begin{tabular}{@{}lS[table-format=5.0]cc@{}}\toprule
        Component & \multicolumn{1}{c}{Error Count} & Errors / \si{\bit} & Cross-Section\tnote{\textdagger} \\\midrule
        Single \cgls{sram} Errors & 18786 & \SI{7.35e-3}{} & \eqmakebox[U][l]{\num{2.77e-08}}\si{\square\cm} \\
        \cgls{tcls} Events & 13 & - & \eqmakebox[U][l]{\num{3.23e-11}}\si{\square\cm} \\
        System Crashes & 0 & - & \eqmakebox[U][l]{\num{5.36e-12}\,}\si{\centi\meter\squared} \\\bottomrule
    \end{tabular}
    \begin{tablenotes}
        \item[\textdagger] Upper-bound with a \SI{95}{\percent} confidence interval \& Poisson distribution. This assumes one additional fault, i.e., device failure after the tests.
    \end{tablenotes}
    \label{tab:neutrons}
\end{threeparttable}
\end{table}

\subsubsection{Memory Errors}

During the neutron exposure, we observed \result{\SI{18786}{errors}} with a median error rate of \result{\SI{492}{\error\per\hour}}.
All of these were single errors detected and corrected by the scrubber that checked the hardware \cgls{ecc} encoding.
Using \Cref{equ:crosssection} allows us to estimate the cross-section of the memory to \result{\SI{2.75(0.02)e-08}{\centi\meter\squared}}.
With \SI{2555904}{\bit}s, we measured an average bit error rate of \result{\SI{1.92e-04}{\error\per\bit\per\hour}} for a cross-section per bit of \result{\SI{1.08(0.01)e-14}{\centi\meter\squared\per\bit}}.
This rate is around \SI{11}{\percent} lower than the findings in~\cite{vargas_radiation_2017}, which evaluated the \SI{14}{\mega\eV} neutron sensitivity of an \cgls{sram} manufactured in the same technology.

\subsubsection{Core Errors}

During the experiment, we measured \result{\SI{13}{}} correctable \cgls{tcls} events.
We estimate the cross-section of recoverable \cgls{tcls} events to be \result{\SI{2.55(0.68)e-11}{\square\cm}}.
No unrecoverable faults were measured. Thus, we can only estimate the upper-bound cross-section of the unrecoverable events to \result{\SI{5.36e-12}{\square\cm}}.

To contextualize, we consider a terrestrial environment to provide estimations for \cgls{mttf}.
Assuming an integral flux density of \SI{20}{\neutron\per\square\cm\per\hour} above \SI{1}{\mega\eV} corresponding to the ground radiation level in New York \cite{gordon_measurement_2004}, the minimum \cgls{mttf} for a terrestrial application with an active \cgls{tcls} recovery mechanism is above \result{\SI{1.06}{}} million years.

\subsection{Proton Tests}

Trikarenos was exposed to a total fluence of \result{\SI{2.91e12}{\proton\per\square\cm}} over \result{\SI{55}{\minute}} of active beam time, a notably high value. This value directly results from our stress testing of the system with the highest flux available in the facility.
We tested a range of beam currents for \SI{200}{\mega\eV} proton with the flux rate extracted from calibration data provided by the facility operator.
\Cref{fig:results} displays the measurements, and \Cref{tab:protons} presents the observed errors alongside estimated cross-sections.

\begin{table}[ht]
\centering
\begin{threeparttable}
    \caption{Cross-section results of Trikarenos under proton beam for a fluence of \result{\SI{2.91e12}{protons\per\centi\meter\squared}}}
    \begin{tabular}{@{}lS[table-format=5.0]cc@{}}\toprule
        Component & \multicolumn{1}{c}{Error count} & Errors / \si{\bit} & Cross-Section\tnote{\textdagger} \\\midrule
        Single \cgls{sram} Errors & 8249 & \SI{3.23e-3}{} & \eqmakebox[U][l]{\num{2.89e-09}}\si{\centi\meter\squared} \\
        \cgls{tcls} Events & 11 & - & \eqmakebox[U][l]{\num{6.76e-12}\,}\si{\square\cm} \\
        System Crashes & 1 & - & \eqmakebox[U][l]{\num{1.91e-12}}\si{\centi\meter\squared} \\\bottomrule
    \end{tabular}
    \begin{tablenotes}
     \item[\textdagger] Upper-bound with a \SI{95}{\percent} confidence interval \& Poisson distribution. This assumes one additional fault, i.e., device failure after the tests.
    \end{tablenotes}
    \label{tab:protons}
\end{threeparttable}
\end{table}

\subsubsection{Memory Errors}

We measured \result{\SI{8249}{errors}} for a median error rate of \result{\SI{9180}{\error\per\hour}}, all corrected by the scrubbers, leading to \cgls{sram} cross-section \result{\SI{2.86(0.03)e-09}{\centi\meter\squared}}.
This results in a cross-section per bit of \result{\SI{1.12(0.01)e-15}{\centi\meter\squared\per\bit}}.

\subsubsection{Core Errors}

During the experiment, we measured \result{\SI{11}{}} recoverable errors and \result{one} unrecoverable error.
Our data indicates that an unresponsive debug module caused the unrecoverable error, an element of the \cgls{soc} currently not protected.
We estimate the cross-section of recoverable \cgls{tcls} events to \result{\SI{5.25(1.51)e-12}{\centi\meter\squared}} and the upper bound for unrecoverable faults to \result{\SI{1.91e-12}{\square\cm}}.
Using this,
we can calculate the minimum \cgls{mftf} of the system with the \cgls{tcls} recovery mechanism to \result{\SI{5.23e+11}{\proton\per\square\cm}}.

\subsubsection{Other Faults}

Trikarenos received a total ionizing dose of more than \result{\SI{1.77e3}{\gray}} throughout the measurements with protons.
No degradation in performance or \cgls{tid}-related errors were observed.
Furthermore, the observed current consumption remained constant within noise-induced bounds, indicating no \cglspl{sel}.

\section{Simulation-based Fault Injection} \label{sec:simulation}

In the simulations, \result{\SI{100000}{faults}} were injected into randomly selected registers of \result{\SI{20921}{\cgls{ff}}} in a randomly selected cycle in \result{\SI{29516}{cycles}} where the main application loop was active, with a single simulation only having one traceability fault.
All registers of the design were injectable and used to check for latent errors, while the \cgls{sram} memory was not injected.
\Cref{tab:fault_injection} shows that over \result{\SI{99}{\percent}} of the injected faults lead to correct application termination.
While most of the correct terminations still included latent errors, most of these latent errors did not affect the remaining \cgls{soc}, indicating that these registers or components are not used in the application tested.

\begin{table}[ht]
\centering
    \caption{Simulation results of fault injection into \cglspl{ff} of Trikarenos}
    \label{tab:fault_injection}
    \begin{tabular}{@{}l@{\hspace{0.5em}}S[table-format=5.0]S[table-format=2.2,table-space-text-post=\si{\percent}]S[table-format=5.0]S[table-format=2.2,table-space-text-post=\si{\percent}]@{}}\toprule
         &  \multicolumn{2}{c}{\textbf{Total}} & \multicolumn{2}{c}{\textbf{In Cores}} \\
        Termination Reason & Amount & Probability & Amount & Probability \\
        \midrule
        \textbf{Correct Termination} & \B 99096 & \B 99.10 \,\si{\percent}  & \B 34524 & \B 100.00\,\si{\percent} \\
        \quad Correct & 18958 & 18.96\,\si{\percent}  & 7642 & 22.14\,\si{\percent} \\
        \quad TCLS & 12283 & 12.28\,\si{\percent}  & 11806 & 34.20\,\si{\percent} \\
        \quad Latent Non-Propagating & 61514 & 61.51\,\si{\percent}  & 13805 & 39.99\,\si{\percent} \\
        \quad Latent Propagating & 6341 & 6.34\,\si{\percent}  & 1271 & 3.68\,\si{\percent} \\
        \textbf{Functional Error} & \B 904 & \B 0.90 \,\si{\percent}  & \B 0 & \B 0.00\,\si{\percent} \\
        \quad Timeout & 822 & 0.82\,\si{\percent}  & 0 & 0.00\,\si{\percent} \\
        \quad Exception & 42 & 0.04\,\si{\percent}  & 0 & 0.00\,\si{\percent} \\
        \quad Incorrect & 40 & 0.04\,\si{\percent}  & 0 & 0.00\,\si{\percent} \\
        \midrule
        \textbf{Total} & \B 100000 & \B 100.00\,\si{\percent} &\B  34524 &\B  100.00\,\si{\percent} \\
        \bottomrule
    \end{tabular}
\end{table}

\hlA{Investigating the confidence of the injection results, based on }\cite{leveugle_statistical_2009}\hlA{, we assumed a normal distribution for high event counts, such as correct terminations (\num{99096}) and functional errors (\num{904}).
Using a conservative probability of ($p = 0.5$), we achieved an absolute margin of error of {±\SI{0.31}{\percent}} with a {\SI{95}{\percent}} confidence interval. 
This corresponds to lower bounds of {\SI{98.79}{\percent}} for correct terminations and an upper bound of {\SI{1.21}{\percent}} for functional errors. 
For rare events like exceptions (\num{42}) and incorrect terminations (\num{40}), where the normal approximation is less suitable, we used the Poisson distribution, resulting in intervals of {\SI{0.030}{\percent}} to {\SI{0.055}{\percent}} and {\SI{0.028}{\percent}} to {\SI{0.053}{\percent}}, respectively.
}
Overall, less than \result{\SI{1}{\percent}} of injections resulted in functional errors.
Furthermore, all injections inside one of the three cores resulted in the correct termination, and the majority either had no effect or were corrected by the \cgls{tcls} mechanism.
Assuming a watchdog timer is used to handle timeouts and exceptions of the system, only \result{\SI{0.04}{\percent}} of errors lead to potentially undetected failures.

\begin{table}[ht]
\centering
    \caption{\hlA{Functional errors of fault injection into \cglspl{ff} of Trikarenos for the different modules.}}
    \label{tab:fault_injection_compoments}
    \begin{tabular}{@{}l@{\hspace{1em}}*4S[table-format=3]cS[table-format=5]}
    \toprule
            & \multicolumn{3}{c}{\textbf{Functional Errors}} & & &  \\
        \multirow{-2}{*}{\textbf{Components}} & {Timeout} & {Exception} & {Incorrect} &  \multicolumn{3}{c}{\multirow{-2}{*}{\textbf{Fault / Injections}}}  \\
        \midrule
            Interconnect & 613 & 22 & 26 & 661 & / & 5176 \\
            Debug Module & 164 & 0 & 1 & 165 & / & 7379 \\
            Peripherals & 26 & 12 & 13 & 51 & / & 41867 \\
            Cores \& Voters & 19 & 2 & 0 & 21 & / & 35934 \\
            Memory & 0 & 6 & 0 & 6 & / & 9644 \\
        \midrule
            \textbf{Total} & \B 822 & \B 42 & \B 40 & \B 904 & \B / & \B 100000 \\
        \bottomrule
    \end{tabular}
\end{table}

\subsubsection{Core Errors}
Of the \result{\SI{100000}{faults}} injected, \result{\SI{34524}{faults}} were injected into the three processor cores' \result{\SI{7194}{\cgls{ff}}}, indicated by the right-most columns in \Cref{tab:fault_injection}.
Of these, none caused a functional error, with \result{\SI{34.20}{\percent}} triggering a \cgls{tcls} recovery event, ensuring correctness.
While latent errors still exist for this configuration, any remaining errors in the cores will be corrected if they cause a divergence in a future operation.
\hlA{Alternatively, a re-synchronization routine can be executed to eliminate all latent errors in the cores. This process involves saving the full state of the cores to memory on the main software stack, resetting the faulty core, and restoring its state from the saved data. The programmer can trigger the routine manually or periodically, effectively mitigating the risk of error accumulation over time.}

Further, the \cgls{tcls} mechanism does contain some \cglspl{ff}, none are directly part of the voters. Thus, any error in the voting can be attributed to interconnect or system errors.

\subsubsection{Functional Errors}
\hlA{A closer look at the remaining functional errors in \Cref{tab:fault_injection_compoments} reveals that }\SI{73}{\percent}\hlA{ of system-level issues are caused by bit-flips in the interconnect, a critical component of the design that currently lacks protection.
Additionally, }\SI{18}{\percent}\hlA{ of errors originate from faults in the JTAG debug module. 
While these primarily affect testing and simulation infrastructure, their impact in a real-world application is likely to be minimal but could still benefit from isolation from the rest of the system.
Additionally, }\SI{6}{\percent}\hlA{ of errors are due to bit-flips in the peripherals.
While only a small portion of the final errors, peripherals can affect the remaining system through interrupt signals or by driving the main memory bus, as is possible with Trikarenos' peripheral subsystem.
Unprotected, disabled components could greatly benefit from proper isolation from the rest of the \cgls{soc}.
Although these peripherals are not heavily utilized in the current tests, their role would become far more critical if Trikarenos were deployed in a complete system, such as a satellite, where latent errors in these modules could pose significant risks.
Finally, }\SI{3}{\percent}\hlA{ of errors are caused by bit-flips inside the cores in the \cgls{odrg} wrapper and the \cgls{ecc} modules.}

\section{Discussion} \label{sec:discussion}

\subsection{Memory Errors}

Although the application running may impact the \cgls{tcls} and total fault rate, the \cgls{sram} fault rate checked by the scrubber should not be affected.
We observe a \result{$>9\times$} higher cross-section of the \cgls{sram} for neutrons when comparing neutron and proton fault rates.
A precise cause for this requires further investigation, but could be attributed to variations in particle types, energy levels, and differences in the devices used across tests.
However, this provides an initial foundation for the validity of neutron beam experiments to evaluate the radiation tolerance of new devices, even in other domains.

We can effectively tune the device's scrub rate using \cgls{sram} cross-sections of \result{\SI{1.08(0.01)e-14}{\centi\meter\squared\per\bit}} for neutrons and \result{\SI{1.12(0.01)e-15}{\centi\meter\squared\per\bit}} for protons.
Trikarenos' scrubber was set to the maximum possible rate for the experiments to ensure all errors were caught without accumulation.
This can be reduced to limit power consumption.
With individual scrubbers for each of the eight memory banks, each scrubber is responsible for 8192 words.
To ensure that the entire memory space is checked between the mean failure rate, the scrubber must complete its memory check every \result{\SI{408}{\milli\second}} for the proton test condition, leading to a word check per bank every \result{\SI{6225}{}} cycles.

\subsection{Core Errors}

In the tested operation mode, Trikarenos detects all corruptions within the cores, both those that may lead to system corruption due to an erroneous branch, corrupted pointer or misdecoded instruction, and data corruption.

Operating in Trikarenos' independent mode, a performance improvement of up to \result{$2.96\times$} with three cores in parallel or a \result{\SI{35}{\percent}} reduction in core power consumption with a single core is possible.
However, in this independent mode, erroneous control flow may lead to stalls that a watchdog timer can catch, or \cgls{sdc} can lead to erroneous data.
Assuming all faults in a core lead to a system crash, the device cross-section in parallel operation increases by at least \result{$3.5\times$} from \result{\SI{2.55(0.68)e-11}{\centi\meter\squared}}
for neutrons and by at least \result{$2.14\times$} from \result{\SI{5.64(1.56)e-12}{\centi\meter\squared}}
for protons.
For single-core operation, assuming an equal distribution of errors in the cores, the cross-section improvement is at least \result{$1.12\times$} from \result{\SI{1.08(0.48)e-11}{\centi\meter\squared}} for neutrons.
Further measurements can help narrow down the improvement, likely showing a significant increase in the reliability factor.
Without \cgls{sram} protection, the improvement is much more significant, even considering that only a subsection of the memory is used and that some data corruption may not cause a system crash.

The \cgls{tcls} recovery routine, requiring around 600 cycles or \SI{4.8}{\micro\second} at \SI{125}{\mega\hertz}, has a negligible impact on performance for the observed error rates.

\subsection{Fault Injection Simulation}
The results presented in Table~\ref{tab:fault_injection} show that all faults injected into the protected part of the \cgls{soc} are corrected by the \cgls{tcls} mechanism, demonstrating its effectiveness.
Additionally, most injections result in latent state mismatches that do not propagate outside the cores or do not affect any other component.
However, these latent errors may affect future computations, where the \cgls{tcls} mechanism will detect and correct them.

\subsection{Comparison of the Experiments and Fault Simulation}
\hlA{The effective average \cgls{ff} cross-section can be estimated by combining experimental and simulation results. This is done by adjusting the observed \cgls{tcls} recoveries using their likelihood from the fault injection simulations, using the following expression:}
\begin{equation}\label{equ:cross_section}
\sigma_{FF} = \frac{\sigma_{TCLS}}{(P_{TCLS} * N_{FF})}
\end{equation} 
where $\sigma_{TCLS}$ is the measured \cgls{tcls} cross-section, $P_{TCLS}$ is the probability that a bit-flip leads to an \cgls{tcls} events and $N_{FF}$ the total number of \cglspl{ff} in Trikarenos.
\hlA{It is important to note that this cross-section reflects the combined effect of raw physical susceptibility and fault masking within the architecture. 
Individual \cglspl{ff} may exhibit different susceptibilities depending on their function and the execution context, and the timing of fault injection can further influence error occurrence. Consequently, this averaged model inherently includes variability that is not captured explicitly in the calculation.}
We estimate an average \cgls{ff} cross-sections of \result{\SI{0.99(0.29)e-14}{\centi\meter\squared\per \cgls{ff}}} for atmospheric neutrons and \result{\SI{2.04(0.64)e-15}{\centi\meter\squared\per \cgls{ff}}} for \SI{200}{\mega\eV} protons.
The neutron value aligns with the measurement reported by \cite{fabero_single_2020} for \SI{14}{\mega\eV}, and the proton value is of the same order of magnitude as that reported by \cite{borghello_single_2023}.
Furthermore, the \cgls{ff} cross-section during neutron experiments is similar to the \cgls{sram} bit-error cross-section, while for proton experiments, the \cgls{ff} cross-section appears higher.
While the error margins for these estimations remain quite large, further consideration may need to be given to the fault injection methods, as only \cgls{ff} errors were considered in this investigation, leaving out \cgls{set} in logic blocks affecting the state.

\subsection{Other}
The neutron and proton experiments also revealed vulnerabilities in the telemetry registers, showing flips of specific bits reflecting immediate, immense increases in measured error counts. While not critical for performance or operations, such registers or similar \cgls{soc} control registers highlight latent errors that do not affect the computation result seen in simulation and may be worth protecting in future designs.

\hlA{Additionally, comparing the functional errors in \Cref{tab:fault_injection_compoments} to the area or the \cgls{ff} count shown in \Cref{fig:trikarenos_area-ff} reveals that vulnerability is not directly correlated with either. 
This observation aligns with the fact that not all components are equally critical or active during operation. 
For instance, while the interconnect occupies a relatively small portion of the system's \cglspl{ff}, it remains a key source of system-level issues due to its critical role in data transfer, which currently lacks protection mechanisms.}

Finally, while unrecoverable errors were very rare in our investigations, it was shown that they are still possible, requiring watchdog circuitry or further protection in future designs. For the Trikarenos \cgls{soc}, their root cause can be traced back to the interconnect, debug module, and peripherals, which would benefit from fault tolerance in a future design.

\section{Conclusion} \label{sec:conclusion}

Exposing the Trikarenos \cgls{soc} to neutrons and protons, we observe most errors affecting the \cgls{sram} for a cross-section per bit of \result{\SI{1.08(0.01)e-14}{\centi\meter\squared\per\bit}} and \result{\SI{1.12(0.01)e-15}{\centi\meter\squared\per\bit}} for atmospheric neutrons and \SI{200}{\MeV} protons, respectively, protected by \cgls{ecc}, thus, not affecting the system's function. Trikarenos' \cgls{tcls} mechanism corrects most of the remaining errors, reducing the cross-section by at least \result{$3.5\times$} from \result{\SI{2.55(0.68)e-11}{\centi\meter\squared}} to below \result{\SI{5.36e-12}{\centi\meter\squared}}, showing \cgls{tcls}' efficacy and Trikarenos' validity for use in critical environments.

Moreover, the \cgls{tcls} mechanism completely protects the cores, preventing all functional errors when bit-flips are injected into the core regions. 
Across the entire \cgls{soc}, only \result{\SI{0.90}{\percent}} of bit-flips result in functional errors, with the majority occurring in the interconnect, one of the critical components of the design that remains unprotected.

\hlA{While this work shows that protecting the largest elements of a microprocessor makes it very robust towards \cglspl{seu}, both induced by particles and in simulation, it further highlights the remaining vulnerabilities within the system. Protecting very active areas such as the interconnect is critical within the design, and enabling proper fault detection and recovery mechanisms or isolation for peripherals merits further research.}
\hlB{Finally, the large number of non-propagating latent errors observed during fault injection simulations, }\result{\SI{61.51}{\percent}}\hlB{ of injected faults, indicates that significant portions of the \cgls{soc} remained inactive in the presented experiments.}
\hlB{Future work is needed to better understand and decrease the number of latent errors remaining in the design, requiring increased observability and a focused software testing methodology, properly considering a wider spectrum of workloads.}

\section*{Acknowledgements}
The authors thank Gianluca Furano from ESA, Thomas Toet from HollandPTC, Edian Annink, and Elijah Cishugi for their support in testing Trikarenos.

\bibliographystyle{IEEEtran}
\bibliography{style,references}

\begin{thebibliography}{10}
\providecommand{\url}[1]{#1}
\csname url@samestyle\endcsname
\providecommand{\newblock}{\relax}
\providecommand{\bibinfo}[2]{#2}
\providecommand{\BIBentrySTDinterwordspacing}{\spaceskip=0pt\relax}
\providecommand{\BIBentryALTinterwordstretchfactor}{4}
\providecommand{\BIBentryALTinterwordspacing}{\spaceskip=\fontdimen2\font plus
\BIBentryALTinterwordstretchfactor\fontdimen3\font minus \fontdimen4\font\relax}
\providecommand{\BIBforeignlanguage}[2]{{%
\expandafter\ifx\csname l@#1\endcsname\relax
\typeout{** WARNING: IEEEtran.bst: No hyphenation pattern has been}%
\typeout{** loaded for the language `#1'. Using the pattern for}%
\typeout{** the default language instead.}%
\else
\language=\csname l@#1\endcsname
\fi
#2}}
\providecommand{\BIBdecl}{\relax}
\BIBdecl

\bibitem{narasimham_hysteresis-based_2012}
\BIBentryALTinterwordspacing
B.~Narasimham \emph{et~al.}, ``A {Hysteresis}-{Based} {D}-{Flip}-{Flop} {Design} in 28 nm {CMOS} for {Improved} {SER} {Hardness} at {Low} {Performance} {Overhead},'' \emph{IEEE Transactions on Nuclear Science}, vol.~59, pp. 2847--2851, Dec. 2012. [Online]. Available: \url{https://ieeexplore.ieee.org/document/6365409}
\BIBentrySTDinterwordspacing

\bibitem{wilson_neutron_2023}
\BIBentryALTinterwordspacing
A.~E. Wilson, M.~Wirthlin, and N.~G. Baker, ``Neutron {Radiation} {Testing} of {RISC}-{V} {TMR} {Soft} {Processors} on {SRAM}-{Based} {FPGAs},'' \emph{IEEE Transactions on Nuclear Science}, vol.~70, pp. 603--610, Apr. 2023. [Online]. Available: \url{https://ieeexplore.ieee.org/document/10012379}
\BIBentrySTDinterwordspacing

\bibitem{neale_neutron_2016}
\BIBentryALTinterwordspacing
A.~Neale and M.~Sachdev, ``Neutron {Radiation} {Induced} {Soft} {Error} {Rates} for an {Adjacent}-{ECC} {Protected} {SRAM} in 28 nm {CMOS},'' \emph{IEEE Transactions on Nuclear Science}, vol.~63, pp. 1912--1917, Jun. 2016. [Online]. Available: \url{https://ieeexplore.ieee.org/document/7497732}
\BIBentrySTDinterwordspacing

\bibitem{di_mascio_open-source_2021}
\BIBentryALTinterwordspacing
S.~Di~Mascio, A.~Menicucci, E.~Gill, G.~Furano, and C.~Monteleone, ``Open-source {IP} cores for space: {A} processor-level perspective on soft errors in the {RISC}-{V} era,'' \emph{Computer Science Review}, vol.~39, Feb. 2021, art. no. 100349. [Online]. Available: \url{https://www.sciencedirect.com/science/article/pii/S1574013720304494}
\BIBentrySTDinterwordspacing

\bibitem{walsemann_strv_2023}
\BIBentryALTinterwordspacing
A.~Walsemann, M.~Karagounis, A.~Stanitzki, and D.~Tutsch, ``{STRV} — a radiation hard {RISC}-{V} microprocessor for high-energy physics applications,'' \emph{Journal of Instrumentation}, vol.~18, Feb. 2023, art. no. C02032. [Online]. Available: \url{https://iopscience.iop.org/article/10.1088/1748-0221/18/02/C02032}
\BIBentrySTDinterwordspacing

\bibitem{de_oliveira_evaluating_2020}
\BIBentryALTinterwordspacing
A.~B. de~Oliveira \emph{et~al.}, ``Evaluating {Soft} {Core} {RISC}-{V} {Processor} in {SRAM}-{Based} {FPGA} {Under} {Radiation} {Effects},'' \emph{IEEE Transactions on Nuclear Science}, vol.~67, pp. 1503--1510, Jul. 2020. [Online]. Available: \url{https://ieeexplore.ieee.org/document/9096401}
\BIBentrySTDinterwordspacing

\bibitem{de_oliveira_evaluating_2023}
\BIBentryALTinterwordspacing
A.~B. de~Oliveira and F.~L. Kastensmidt, ``Evaluating {Fault}-{Tolerant} {Techniques} on {COTS} {RISC}-{V} {NOEL}-{V} {Processor} in {Zynq} {UltraScale}+ {FPGA} {Under} {Proton} {Testing},'' \emph{IEEE Transactions on Nuclear Science}, vol.~70, pp. 1708--1715, Aug. 2023. [Online]. Available: \url{https://ieeexplore.ieee.org/document/10138560}
\BIBentrySTDinterwordspacing

\bibitem{santos_hybrid_2024}
\BIBentryALTinterwordspacing
D.~A. Santos \emph{et~al.}, ``Hybrid {Hardening} {Approach} for a {Fault}-{Tolerant} {RISC}-{V} {System}-{On}-{Chip},'' \emph{IEEE Transactions on Nuclear Science}, vol.~71, pp. 1722--1730, Aug. 2024. [Online]. Available: \url{https://ieeexplore.ieee.org/document/10540019}
\BIBentrySTDinterwordspacing

\bibitem{rogenmoser_trikarenos_2023}
\BIBentryALTinterwordspacing
M.~Rogenmoser and L.~Benini, ``Trikarenos: {A} {Fault}-{Tolerant} {RISC}-{V}-based {Microcontroller} for {CubeSats} in 28nm,'' in \emph{{IEEE} {International} {Conference} on {Electronics}, {Circuits} and {Systems} ({ICECS})}, Istanbul, Turkey, Dec. 2023, pp. 662--665. [Online]. Available: \url{https://ieeexplore.ieee.org/document/10382727}
\BIBentrySTDinterwordspacing

\bibitem{ecss_secretariat_space_2023}
\BIBentryALTinterwordspacing
{ECSS Secretariat}, ``Space engineering: {Engineering} techniques for radiation effects mitigation in {ASICs} and {FPGAs} handbook,'' ESA Requirements and Standards Section, Noordwijk, The Netherlands, Handbook ECSS-E-HB-20-40A, Oct. 2023. [Online]. Available: \url{https://amstel.estec.esa.int/tecedm/website/docs_generic/ECSS-E-HB-20-40A(11October2023).pdf}
\BIBentrySTDinterwordspacing

\bibitem{rogenmoser_hybrid_2025}
\BIBentryALTinterwordspacing
M.~Rogenmoser, Y.~Tortorella, D.~Rossi, F.~Conti, and L.~Benini, ``Hybrid {Modular} {Redundancy}: {Exploring} {Modular} {Redundancy} {Approaches} in {RISC}-{V} {Multi}-core {Computing} {Clusters} for {Reliable} {Processing} in {Space},'' \emph{ACM Trans. Cyber-Phys. Syst.}, vol.~9, pp. 8:1--8:29, Jan. 2025. [Online]. Available: \url{https://dl.acm.org/doi/10.1145/3635161}
\BIBentrySTDinterwordspacing

\bibitem{leibson_nasa_2023}
\BIBentryALTinterwordspacing
S.~Leibson, ``\BIBforeignlanguage{en-US}{{NASA} {Recruits} {Microchip}, {SiFive}, and {RISC}-{V} to {Develop} 12-{Core} {Processor} {SoC} for {Autonomous} {Space} {Missions}},'' \emph{\BIBforeignlanguage{en-US}{EEJournal}}, Jan. 2023. [Online]. Available: \url{https://www.eejournal.com/article/nasa-recruits-microchip-sifive-and-risc-v-to-develop-12-core-processor-soc-for-autonomous-space-missions/}
\BIBentrySTDinterwordspacing

\bibitem{schiavone_quentin_2018}
\BIBentryALTinterwordspacing
P.~D. Schiavone, D.~Rossi, A.~Pullini, A.~Di~Mauro, F.~Conti, and L.~Benini, ``Quentin: an {Ultra}-{Low}-{Power} {PULPissimo} {SoC} in 22nm {FDX},'' in \emph{2018 {IEEE} {SOI}-{3D}-{Subthreshold} {Microelectronics} {Technology} {Unified} {Conference} ({S3S})}, Burlingame, California, USA, Oct. 2018, pp. 20--22. [Online]. Available: \url{https://ieeexplore.ieee.org/document/10382727}
\BIBentrySTDinterwordspacing

\bibitem{rea_powerpc_2005}
D.~Rea, D.~Bayles, P.~Kapcio, S.~Doyle, and D.~Stanley, ``{PowerPC} ™ {RAD750} ™ - {A} {Microprocessor} for {Now} and the {Future},'' in \emph{2005 {IEEE} {Aerospace} {Conference}}, Big Sky, Montana, USA, Mar. 2005, pp. 2390--2394.

\bibitem{berger_quad-core_2015}
R.~Berger \emph{et~al.}, ``Quad-core radiation-hardened system-on-chip power architecture processor,'' in \emph{2015 {IEEE} {Aerospace} {Conference}}, Big Sky, Montana, USA, Mar. 2015, pp. 2250--2261.

\bibitem{johansson_rad-hard_2016}
\BIBentryALTinterwordspacing
F.~Johansson \emph{et~al.}, ``Rad-{Hard} {Microcontroller} {For} {Space} {Applications},'' in \emph{{AMICSA} \& {DSP} 2016 {Proceedings}}.\hskip 1em plus 0.5em minus 0.4em\relax Gothenburg, Sweden: ESA, Jun. 2016, pp. 3--7. [Online]. Available: \url{https://indico.esa.int/event/102/attachments/22/27/Proceedings_2016_AMICSA.pdf}
\BIBentrySTDinterwordspacing

\bibitem{andersson_gaisler_2022}
\BIBentryALTinterwordspacing
J.~Andersson, ``Gaisler {NOEL}-{V} {SoC} {Applications} and {Ecosystem},'' ESA ESTEC, Noordwijk, Netherlands, Dec. 2022. [Online]. Available: \url{http://microelectronics.esa.int/riscv/rvws2022/presentations/06_ESA_RISC-V_in_Space-NOEL-V.pdf}
\BIBentrySTDinterwordspacing

\bibitem{schiavone_slow_2017}
\BIBentryALTinterwordspacing
D.~P. Schiavone \emph{et~al.}, ``Slow and steady wins the race? {A} comparison of ultra-low-power {RISC}-{V} cores for {Internet}-of-{Things} applications,'' in \emph{2017 27th {International} {Symposium} on {Power} and {Timing} {Modeling}, {Optimization} and {Simulation} ({PATMOS})}, Thessaloniki, Greece, Sep. 2017, pp. 185--192. [Online]. Available: \url{https://ieeexplore.ieee.org/document/8106976}
\BIBentrySTDinterwordspacing

\bibitem{borghello_total_2023}
\BIBentryALTinterwordspacing
G.~Borghello \emph{et~al.}, ``Total ionizing dose effects on ring-oscillators and {SRAMs} in a commercial 28 nm {CMOS} technology,'' \emph{Journal of Instrumentation}, vol.~18, Feb. 2023, art. no. C02003. [Online]. Available: \url{https://dx.doi.org/10.1088/1748-0221/18/02/C02003}
\BIBentrySTDinterwordspacing

\bibitem{cazzaniga_progress_2018}
\BIBentryALTinterwordspacing
C.~Cazzaniga and C.~D. Frost, ``Progress of the {Scientific} {Commissioning} of a fast neutron beamline for {Chip} {Irradiation},'' \emph{Journal of Physics: Conference Series}, vol. 1021, May 2018, art. no. 012037. [Online]. Available: \url{https://iopscience.iop.org/article/10.1088/1742-6596/1021/1/012037}
\BIBentrySTDinterwordspacing

\bibitem{rovituso_characterisation_2023}
\BIBentryALTinterwordspacing
M.~Rovituso \emph{et~al.}, ``Characterisation of the {HollandPTC} {R}\&{D} proton beamline for physics and radiobiology studies,'' \emph{arXiv.org}, Feb. 2023. [Online]. Available: \url{https://arxiv.org/abs/2302.09943v1}
\BIBentrySTDinterwordspacing

\bibitem{johnson_poisson_2005}
\BIBentryALTinterwordspacing
N.~L. Johnson, A.~W. Kemp, and S.~Kotz, ``Poisson {Distribution},'' in \emph{Univariate {Discrete} {Distributions}}.\hskip 1em plus 0.5em minus 0.4em\relax John Wiley \& Sons, Ltd, 2005, p. 176. [Online]. Available: \url{https://onlinelibrary.wiley.com/doi/abs/10.1002/0471715816.ch4}
\BIBentrySTDinterwordspacing

\bibitem{vargas_radiation_2017}
\BIBentryALTinterwordspacing
V.~Vargas \emph{et~al.}, ``Radiation {Experiments} on a 28 nm {Single}-{Chip} {Many}-{Core} {Processor} and {SEU} {Error}-{Rate} {Prediction},'' \emph{IEEE Transactions on Nuclear Science}, vol.~64, pp. 483--490, Jan. 2017. [Online]. Available: \url{https://ieeexplore.ieee.org/document/7779088}
\BIBentrySTDinterwordspacing

\bibitem{gordon_measurement_2004}
\BIBentryALTinterwordspacing
M.~Gordon \emph{et~al.}, ``Measurement of the flux and energy spectrum of cosmic-ray induced neutrons on the ground,'' \emph{IEEE Transactions on Nuclear Science}, vol.~51, pp. 3427--3434, Dec. 2004. [Online]. Available: \url{http://ieeexplore.ieee.org/document/1369506/}
\BIBentrySTDinterwordspacing

\bibitem{leveugle_statistical_2009}
\BIBentryALTinterwordspacing
R.~Leveugle, A.~Calvez, P.~Maistri, and P.~Vanhauwaert, ``Statistical fault injection: {Quantified} error and confidence,'' in \emph{Automation \& {Test} in {Europe} {Conference} \& {Exhibition} 2009 {Design}}, Nice, France, Apr. 2009, pp. 502--506, iSSN: 1558-1101. [Online]. Available: \url{https://ieeexplore.ieee.org/document/5090716}
\BIBentrySTDinterwordspacing

\bibitem{fabero_single_2020}
\BIBentryALTinterwordspacing
J.~C. Fabero \emph{et~al.}, ``Single {Event} {Upsets} {Under} 14-{MeV} {Neutrons} in a 28-nm {SRAM}-{Based} {FPGA} in {Static} {Mode},'' \emph{IEEE Transactions on Nuclear Science}, vol.~67, pp. 1461--1469, Jul. 2020. [Online]. Available: \url{https://ieeexplore.ieee.org/document/9020124}
\BIBentrySTDinterwordspacing

\bibitem{borghello_single_2023}
\BIBentryALTinterwordspacing
G.~Borghello, D.~Ceresa, R.~Pejasinovic, F.~Piernas~Diaz, G.~Bergamin, and K.~Kloukinas, ``\BIBforeignlanguage{en}{Single {Event} {Effects} characterization of a commercial 28 nm {CMOS} technology},'' in \emph{\BIBforeignlanguage{en}{Topical {Workshop} on {Electronics} for {Particle} {Physics} ({TWEPP})}}.\hskip 1em plus 0.5em minus 0.4em\relax Geremeas: CERN, Oct. 2023. [Online]. Available: \url{https://indico.cern.ch/event/1255624/contributions/5443894/}
\BIBentrySTDinterwordspacing

\end{thebibliography}

\end{document}